\documentclass[10pt,journal,final]{article}
\usepackage[margin=1.4in]{geometry}  
\usepackage{graphicx}
\usepackage{lipsum} 

\usepackage[utf8]{inputenc}
\usepackage{xspace,empheq,fancybox,amsmath,amssymb,amsfonts,epstopdf,epsfig,syntonly,times,amsthm} 
\usepackage{psfrag,color,bm,array,tabularx, subfigure}
\usepackage{cite,url,footnote,xspace,syntonly, algorithmic}
\usepackage{verbatim,multirow}
\usepackage[linesnumbered,ruled,vlined]{algorithm2e}
\usepackage[T1]{fontenc}
\usepackage{textcomp}
\usepackage{xcolor}
\usepackage{mathtools}
\usepackage{titling}
\usepackage{makecell}
\usepackage{bbm, booktabs, makecell, subcaption, graphicx, titlesec,adjustbox,lipsum}

\usepackage{cuted}
\usepackage[caption=false,font=footnotesize]{subfig}
\usepackage{nomencl}
\makenomenclature
\newcolumntype{M}[1]{>{\centering\arraybackslash}m{#1}}
\newcolumntype{N}{@{}m{0pt}@{}}

\def\BibTeX{{\rm B\kern-.05em{\sc i\kern-.025em b}\kern-.08em
    T\kern-.1667em\lower.7ex\hbox{E}\kern-.125emX}}

\newcommand{\Pfdroop}{\text{$P$-\,$\omega$ }}
\newcommand{\QVdroop}{\text{$Q$-$V$ }}
\usepackage{tikz}

\newtheorem{remark}{\bfseries Remark}

\input{mysymbol.sty}
\allowdisplaybreaks

\pretitle{
  \begin{center}
    \LARGE\bfseries
}
\posttitle{%
  \end{center}%
  \vskip 0.5em
}

\preauthor{%
  \begin{center}%
    \normalsize
}
\postauthor{%
  \end{center}%
  \vskip 1em
}

\predate{}  
\postdate{}

\title{%
  Physics-Informed Learning of Proprietary Inverter \\
  Models for Grid Dynamic Studies
}

\author{%
  Kyung-Bin Kwon, Sayak Mukherjee, Ramij R. Hossain, Marcelo Elizondo%
  \thanks{Authors are with the Pacific Northwest National Laboratory,
    Richland, WA 99352, USA. This research is supported by the
    E‑COMP Initiative at PNNL (Contract DE‑AC05‑76RL01830).}%
}

\date{}

\begin{document}
\maketitle

%

\begin{abstract}
This letter develops a novel physics-informed neural ordinary differential equations-based framework to emulate the proprietary dynamics of the inverters --- essential for improved accuracy in grid dynamic simulations.
In current industry practice, the original equipment manufacturers (OEMs) often do not disclose the exact internal controls and parameters of the inverters, posing significant challenges in performing accurate dynamic simulations and other relevant studies, such as gain tunings for stability analysis and controls. To address this, we propose a Physics-Informed Latent Neural ODE Model (PI-LNM) that integrates system physics with neural learning layers to capture the unmodeled behaviors of proprietary units. The proposed method is validated using a grid-forming inverter (GFM) case study, demonstrating improved dynamic simulation accuracy over approaches that rely solely on data-driven learning without physics-based guidance.
\end{abstract}


%

\section{Introduction} \label{sec:int}
Power systems dynamical analysis and studies depend heavily on the precision and accuracy of the simulation models utilized to represent various grid assets and components.
With the increasing penetration of inverter-based resources (IBRs), replicating the inverter dynamics has become essential to represent grid behaviors observed in practice. 
Traditionally, power system simulators incorporate well-validated detailed representations of conventional generation assets, e.g., synchronous generators (SGs). There are recent efforts to develop generic models of inverters. 
however, the internal control architectures and parameters of these inverters are often unknown or inaccessible to operators as OEMs often restrict the proprietary information related to those models \cite{kroposki2024unifi}. This poses significant challenges to develop high-precision dynamic models needed for system-level studies. Recent advances in AI and deep learning have demonstrated the potential of learning-based latent dynamical systems capturing complex nonlinear interactions in physical systems \cite{goodfellow2014generative}. These approaches can be further enhanced by integrating nominal physics knowledge in the learning process, as explored in prior works \cite{raissi2019physics}.

To this end, we propose a \textit{physics-informed neural latent ordinary differential equation modeling framework} to infer and emulate the dynamic behavior of proprietary inverter models where nominal inverter dynamics are explicitly embedded along with additional neural learning layers to mimic the original dynamics. 
The main contribution of this letter is summarized as follows: The proposed architecture leverages a baseline representative inverter control structure to inject partial physical priors into the learning process, resulting in a hybrid model that combines neural network components with physics-informed dynamics. We introduce a physics-informed latent neural ordinary differential equation-based model (PI-LNM) framework that blends data-driven learning with structural knowledge from inverter dynamics to replicate the behavior of black-box proprietary models.

\section{Proposed Methodology} \label{sec:ODE}
Let us consider the power grid composed of $M$ buses, with $M_m$ being the number of resource buses with either SGs or GFM-based resources. The dynamics of the SGs are well-established \cite{10556716} and can be compactly represented as, $\dot{x}_s = f_{\mbox{sync}} (x_s, V)$, where $f_{\mbox{sync}}(.)$ presents the functional form of the dynamics with $V = [V_{1_{Re}}, \dots, V_{M_{Re}}, V_{1_{Im}}, \dots, V_{M_{Im}}]$, and $x_s$ denotes the SG states. 
Without loss of generality, in this paper, we study proprietary dynamic models of the GFMs --- compactly denoted as:
    $\dot{x}_f = f_{\mbox{gfm}} (x_f,  V),$ where GFM state vectors are stacked together as $x_f$. 
Here, we assume that system operators have some knowledge of the generic GFM models, for instance, the dynamic representation of the WECC-approved REGFM\_A1 model \cite{du2023model} with \Pfdroop and \QVdroop droop. However, the exact model provided by the OEM is unknown. Therefore, our objective is to approximate the dynamics of the actual GFM model with a data-driven learning-based neural model infusing the generic physics knowledge, represented in \eqref{gfm}.
\begin{align}\label{gfm}
    \dot{x}_f = N_\theta(\hat{f}_{\mbox{gfm}} (x_f,  V)).
\end{align}
where $N_\theta(.)$ represents a learnable neural function, $\hat{f}_{\mbox{gfm}} (.)$ considers generic physics of GFM model, e.g. the dynamics described in \cite{du2023model}. 
This letter specifically investigates the physics-informed latent ODE framework to replicate the behavior of proprietary GFM models. First, we briefly introduce the neural latent ODE (NODE) models. Next, the learning framework utilized for GFM models will be discussed.

\subsection{Physics-informed Latent Neural ODE}
Our approach combines NODE models that capture unknown characteristics from real-world field data with physical information derived from known mathematical dynamics \cite{moya2023approximating}. 
The process of a latent ODE model can be formalized as follows \cite{rubanova2019latent}. The initial latent state \( z_0 \) is sampled from a prior distribution $p(.)$: $z_0 \sim p(z_0).$
In the standard approach, the latent trajectory is computed by solving an ODE parameterized by a neural network function \( f_\theta \) over the observation times \( t_0, \dots, t_N \):
\begin{equation}
z_0, z_1, \dots, z_N = \text{ODESolve}(f_\theta, z_0, (t_0, t_1, \dots, t_N)).
\end{equation}
The function \( f_\theta \) models the time derivative of the latent state as
$
\frac{d z}{d t} = f_\theta(z, t),
$
where \( \theta \) represents trainable parameters. We propose a PI-LNM where some representative dynamic equations are considered (e.g., $\hat{f}_{\text{gfm}}(.)$) along with additional neural ODEs:
\begin{equation}
\label{eq:hybrid_ode}
\frac{d}{d t} \begin{bmatrix}
     z \\ z_r
\end{bmatrix}
    = \begin{bmatrix}
        f_\theta(z, z_r, t) \\ \hat{f}_{\text{gfm}}(z, z_r, t)
    \end{bmatrix}
\end{equation}
where the dynamics of $z_r$ uses a representative model:
\begin{equation}
\frac{d z_{r}}{d t} = \hat{f}_{\text{gfm}}(z_{r},z, t), \label{eq:known}
\end{equation}
The decoder function, a trainable neural network, maps the latent states back to the observation space $x_i \sim p(x_i | z_i), \quad i = 0, \dots, N$. In standard designs, this can be a multi-layer perceptron (MLP) based neural network.  

To estimate the posterior distribution over latent trajectories, an ODE-RNN was proposed as an encoder \cite{chen2018neural}. This approach entails encoding the sequence of observations \( \{x_i, t_i\}_{i=0}^N \) using an encoder neural ODE and an RNN to produce a hidden state representation at the final time step, denoted as \( h_T \). A neural network \( g \) maps this hidden state to the parameters of the approximate posterior distribution:
\begin{equation*}
q_\phi(z_0 | \{x_i, t_i\}_{i=0}^N) = \mathcal{N}(\mu_{z_0}, \sigma_{z_0}) \text{ where } (\mu_{z_0}, \sigma_{z_0}) = g(h_T).
\end{equation*}

Training a latent ODE model involves jointly optimizing the parameters of the encoder (ODE-RNN) and the decoder, and the latent ODE's neural functions \( f_\theta \) by maximizing the evidence lower bound (ELBO):
\begin{align}
\label{eq:elbo}
\mathcal{L}_{ELBO} &= \mathbb{E}_{q_\phi(z_0 | \{x_i, t_i\})} \Bigg[ \sum_{i=0}^N \log p(x_i | z_i) \Bigg] - D_{KL}(q_\phi(z_0 | \{x_i, t_i\}) \,\|\, p(z_0)) 
\end{align}

\begin{remark}
    Since latent ODEs are a generative model in which ODEs govern the dynamics of latent variables \cite{rubanova2019latent}, our approach merges the generative representation power of latent models with the infusion of physical information. As such, latent ODEs differ from traditional autoregressive models such as RNNs and their variants (e.g., GRU and LSTM).
\end{remark}
\subsection{Learning Framework for GFM models}

\begin{figure}[h]
    \centering
    \includegraphics[width=0.8\linewidth]{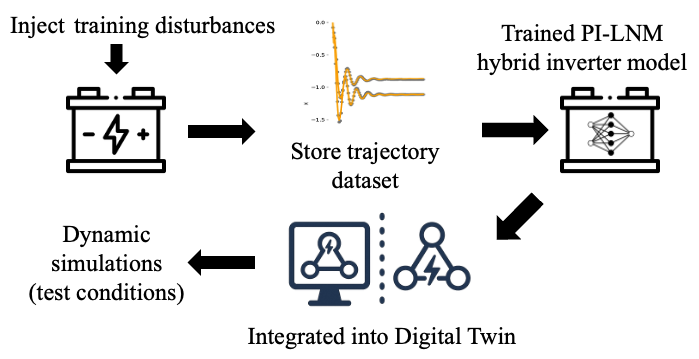}
    \caption{Schematic overview of the learning framework}
    \label{fig:overview}
\end{figure}

To implement the physics-informed latent ODE model that mimics the GFM dynamics, we consider the following steps, also depicted in Fig. \ref{fig:overview}: 

\noindent \textit{Step 1:} In a controlled experimental environment, the operator first interconnects the proprietary actual GFM unit to a source (replicating an infinite bus), along with a load connected at the POI. If the experimental setup is not feasible, the black-box proprietary software models can be connected in similar fashion to a grid simulator. \\
\textit{Step 2:} Generate different events by changing the active power consumed by the load in a certain large range, and store $T$ number of time-steps. We store the trajectories of the state tuple $\mathbf{x}$ for $K$ number of trajectories. \\
\textit{Step 3:} Process the observation data in batches through the encoder network to estimate an initial condition and perform roll-out forward in time using the physics-informed Neural ODE function \eqref{eq:hybrid_ode}.  \\
\textit{Step 4:} Using the optimization framework as described in equation \eqref{eq:elbo}, we compute ELBO losses and perform back-prop to learn the latent ODE model where $|z| >> |x|$ ($|.|$ denotes cardinality).\\
\textit{Step 5:} Once the neural GFM dynamics is learned, we utilize this as our equivalent model with physics-enabled prior embedded for simulations. To this end, we connect the learned dynamical block to the target grid simulation platform with other grid components and perform dynamic simulations. 

\begin{remark}
    As shown in Fig.~\ref{fig:overview}, our proposed model can be directly incorporated into the digital twin framework because it learns a compact, physics-informed representation of GFM dynamics from field data. This enables seamless coupling with existing grid simulators, offering a real-time, data-driven surrogate that preserves underlying physical laws while significantly reducing computational complexity.
\end{remark}

%

\section{Simulation Results}
To mimic an actual power grid setup, we use an emulator model with an assumed proprietary restricted GFM unit and a test system to generate data and test the performance of the learned PI-LNM model. To generate the training data, at $t=0$, we introduce a change in the load between $0.5$ p.u. and $5$ p.u. and record a trajectory of the observations $\mathbf{x}:= [\theta_j, \omega_j, V^e_j, V_j, P^n_j, Q^n_j]$, which includes both the GFM states (angles, frequencies, voltage control errors, and the internal voltage magnitude) and the active, reactive power delivered through the network. The time interval is set to $\Delta t=0.01$ seconds, and we observe for 10 seconds. 
%
\begin{figure}[htbp]
    \centering
    \includegraphics[width=0.7\linewidth]{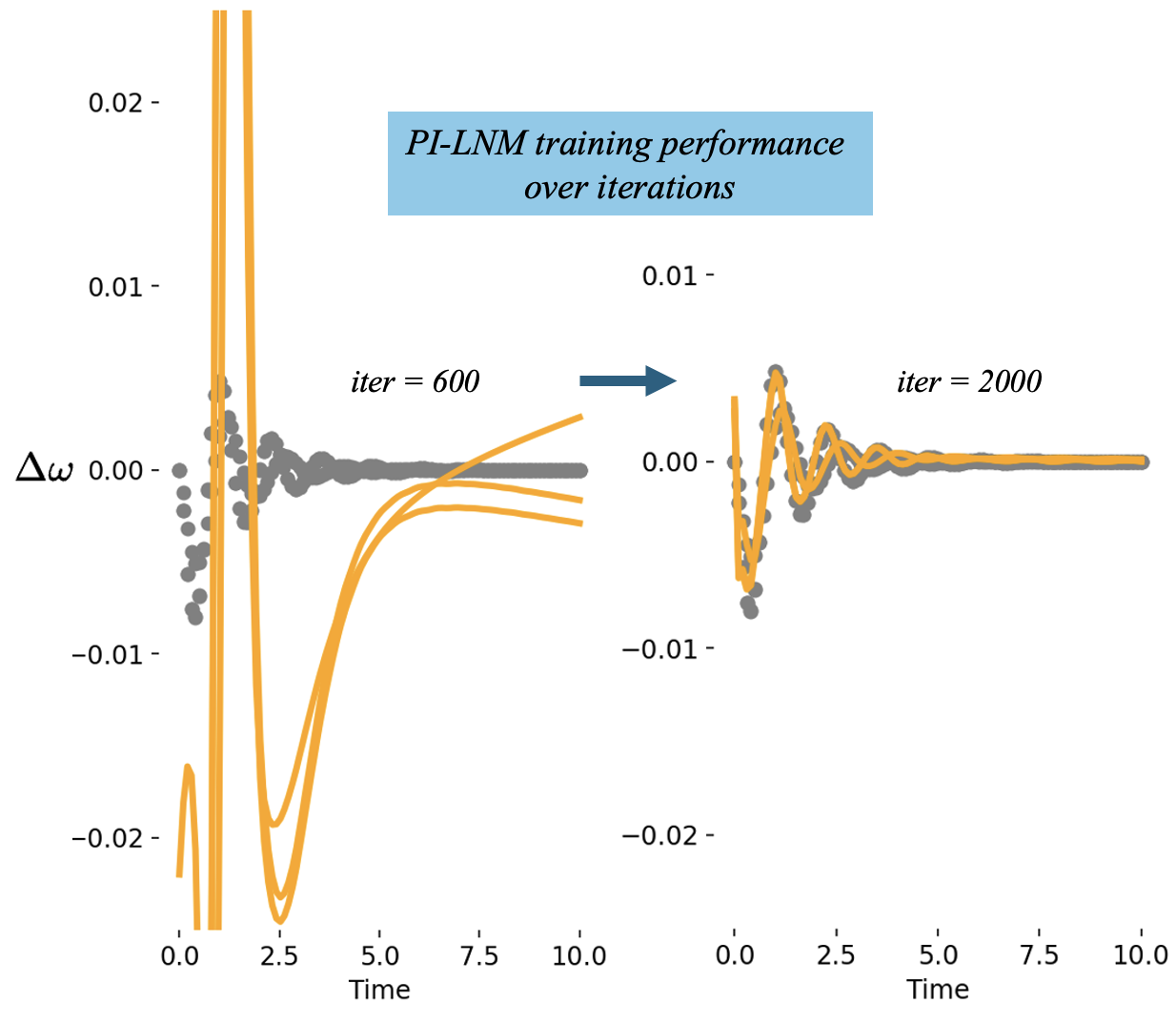}
    \caption{PI-LNM training performance}
    \label{fig:freq_training}
    \centering
    \includegraphics[width=0.7\linewidth]{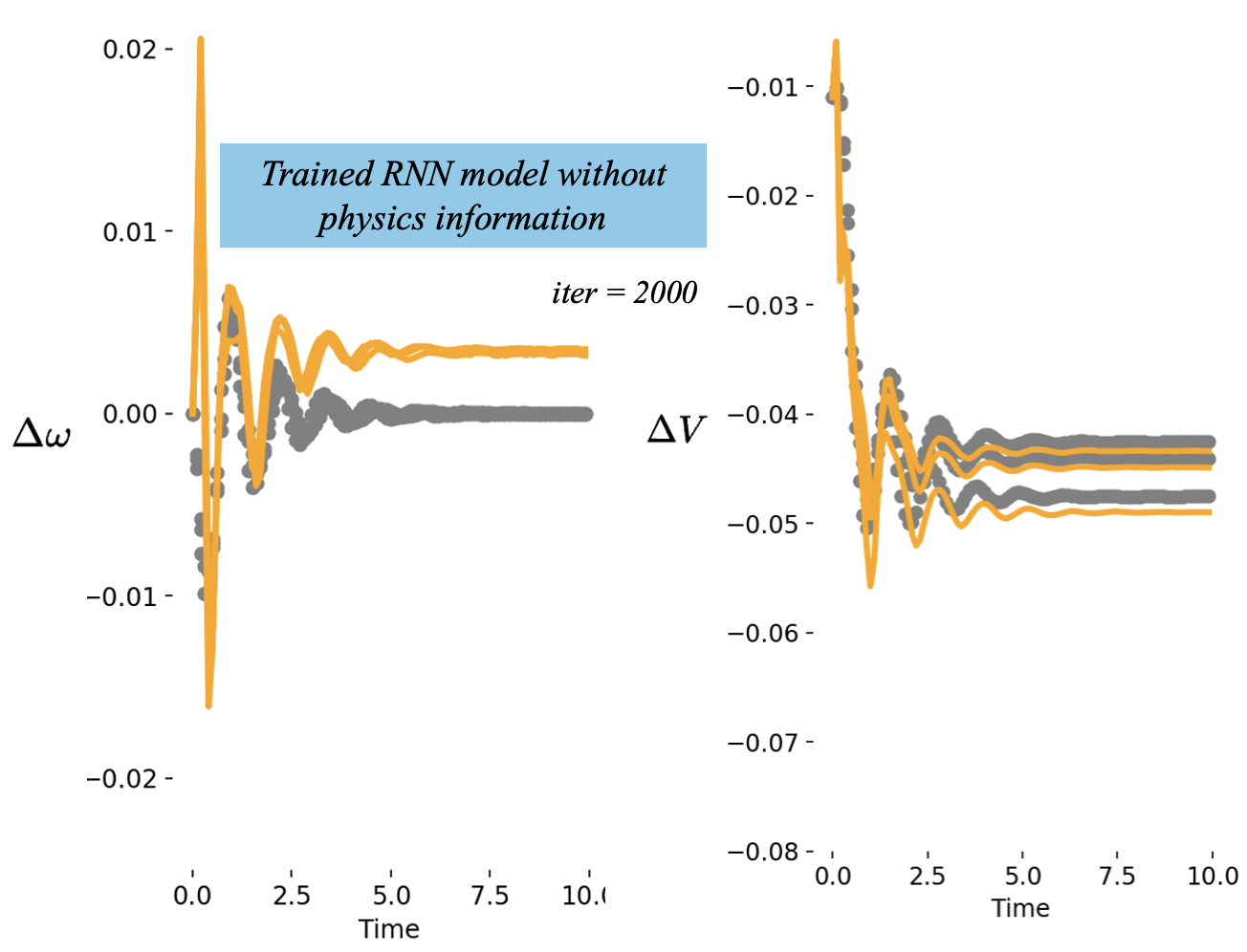}
    \caption{Insufficient training performance of RNN-based (not physics-informed) modeling framework}
    \label{fig:rnn}
\end{figure}
We generate 2000 samples, each consisting of six trajectories of state variables. Using this data, we apply latent ODE to learn the GFM dynamics. A total of 24 latent variables are set,
out of which the first four variables correspond to $[\theta_j, \omega_j, V^e_j, V_j]$.
Here, we assume that the exact values of the GFM parameters $\hat{f}_{GFM}$ are unknown, and introduce a $\pm 20\%$ error from the approximated GFM droop and gain parameters $\{m_p, m_q, k_{pv}, k_{iv}\}$. 
We utilize these approximated parameters in $\hat{f}_{GFM}$, whereas the training data is generated with the actual dynamics. 
The batch size is set to $200$ out of the $2000$ samples, with an initial learning rate of $0.02$. The number of iterations is set at $2000$. To show the effectiveness of our latent-ODE-based model, we compare three cases: \textit{Actual GFM dynamics}, \textit{RNN-based without physics information}, and \textit{PI-LNM}. 
\begin{figure}[t]
    \centering
\includegraphics[width=0.8\linewidth]{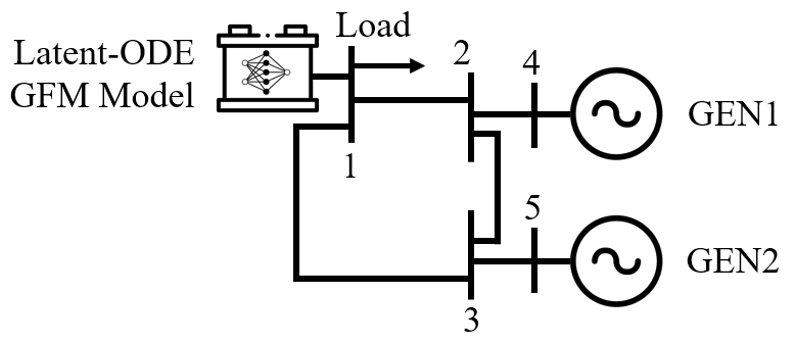}
    \caption{Trained PI-LNM GFM model integrated to simulation}
    \label{fig:test}
    \end{figure}
    \begin{figure}[t]
    \centering
    \includegraphics[width=\linewidth]{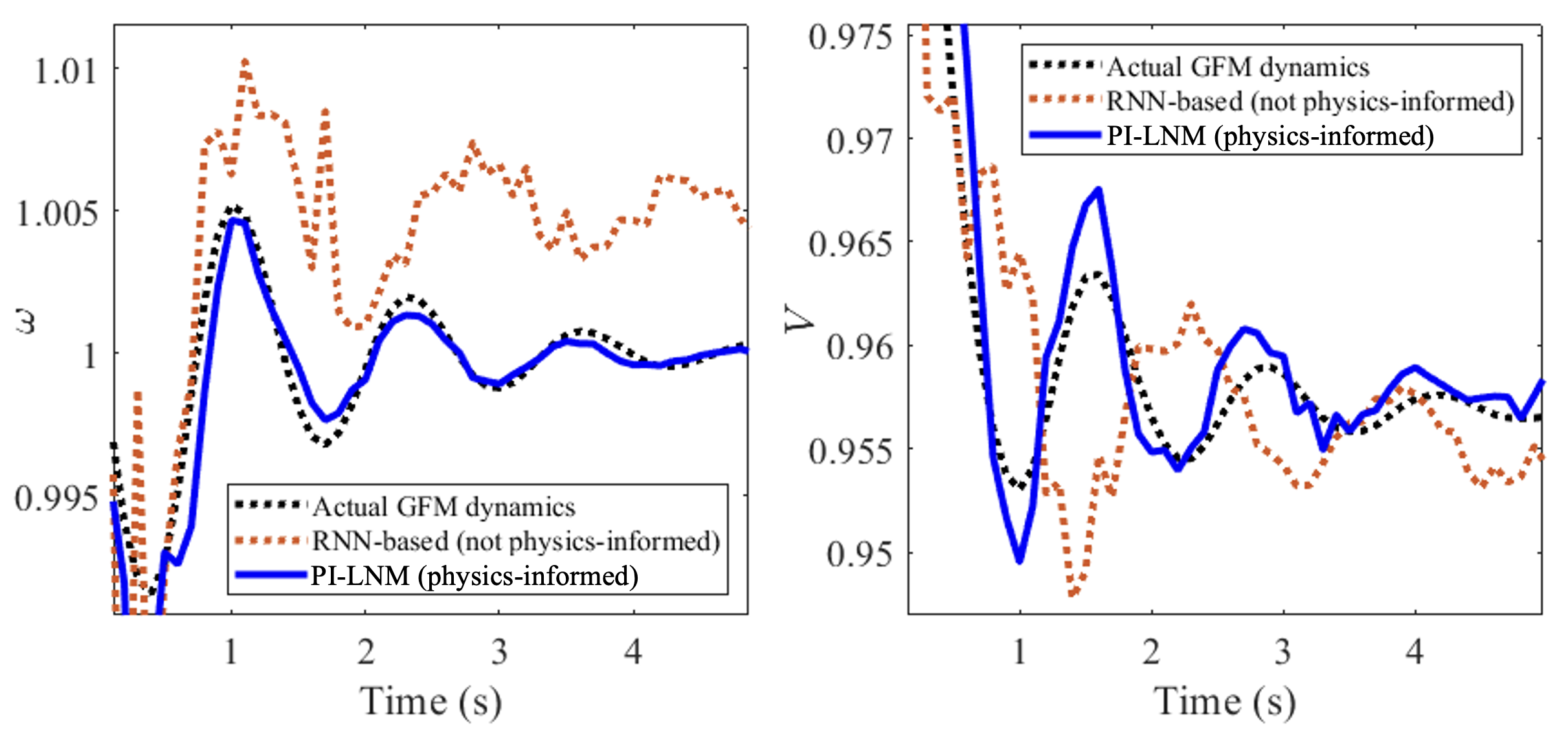}
    \caption{Test-time performance evaluation}
    \label{fig:perf}
\end{figure}
The training results of \textit{LNM} are shown in Fig.~\ref{fig:freq_training}, illustrating the results at iterations $600$, and $2000$, respectively. As seen in the figure, with each iteration, the latent ODE agent progressively estimates the time-series trajectory with increasing accuracy. In addition, the training trajectory of \textit{RNN}-based model is depicted in Fig.~\ref{fig:rnn}, showing at iteration $2000$, and compared to \textit{LNM}, \textit{RNN} cannot converge to the target due to a lack of physics guidance.

With the trained model, we conduct tests on an experimental test system, as shown in Fig.~\ref{fig:test}. Similar to the training setup, the load is connected at bus 1, where the GFM is located, and sudden, randomly chosen changes in load occur at $t=0$. We observe the trajectory for $5$ seconds. 
With an initial power flow solution, the trajectory is accumulated using estimates of the next steps through the trained NODE agent, based on the previous observations. Fig.~\ref{fig:perf} shows the trajectory comparison where the proposed model could effectively mimic the dynamics of the actual GFM with respect to a standard RNN-based model.

\begin{table}[h!]
\centering
\caption{Comparison of RMSE values for Prediction}
\begin{tabular}{c||c||c||c}
\hline
\textbf{Metric (RMSE)} & \textbf{RNN} & \textbf{PI-LNM} & \textbf{Improvement (\%)} \\
\hline

Voltage (pu)    & $6.0\times10^{-3}$                 & $4.0\times10^{-3}$                  & $33.3\%$                    \\
Frequency (Hz)  & $5.8\times10^{-3}$                   & $2.2\times10^{-3}$                   & $62.1\%$                  \\
\hline
\end{tabular}
\label{tab:rmse_comparison}
\end{table}

\section{Conclusion}
In this paper, we developed a PI‐LNM framework to emulate proprietary GFM dynamics by embedding nominal droop‐control priors into a neural‐ODE architecture. The proposed PI‐LNM combines an ODE‐RNN encoder with a physics‐augmented latent ODE decoder to learn both known and unmodeled behaviors from time‐series data. Numerical experiments show that PI‐LNM outperforms an RNN baseline, reducing RMSE in voltage and frequency trajectories by over 30\% and 60\%, respectively. This approach demonstrates that incorporating physics enhances learning efficiency and generalization, offering a practical pathway for emulation of black‐box inverter models in power system analysis.

\bibliography{ref}

\begin{thebibliography}{1}
\providecommand{\url}[1]{#1}
\csname url@samestyle\endcsname
\providecommand{\newblock}{\relax}
\providecommand{\bibinfo}[2]{#2}
\providecommand{\BIBentrySTDinterwordspacing}{\spaceskip=0pt\relax}
\providecommand{\BIBentryALTinterwordstretchfactor}{4}
\providecommand{\BIBentryALTinterwordspacing}{\spaceskip=\fontdimen2\font plus
\BIBentryALTinterwordstretchfactor\fontdimen3\font minus \fontdimen4\font\relax}
\providecommand{\BIBforeignlanguage}[2]{{%
\expandafter\ifx\csname l@#1\endcsname\relax
\typeout{** WARNING: IEEEtran.bst: No hyphenation pattern has been}%
\typeout{** loaded for the language `#1'. Using the pattern for}%
\typeout{** the default language instead.}%
\else
\language=\csname l@#1\endcsname
\fi
#2}}
\providecommand{\BIBdecl}{\relax}
\BIBdecl

\bibitem{kroposki2024unifi}
B.~Kroposki \emph{et~al.}, ``{UNIFI specifications for grid-forming inverter-based resources (Version 2)},'' National Renewable Energy Laboratory (NREL), Golden, CO, USA, Technical Report NREL/TP-5D00-89269, 2024.

\bibitem{goodfellow2014generative}
I.~Goodfellow, J.~Pouget-Abadie, M.~Mirza, B.~Xu, D.~Warde-Farley, S.~Ozair, A.~Courville, and Y.~Bengio, ``{Generative adversarial nets},'' in \emph{Advances in Neural Information Processing Systems}, 2014, pp. 2672--2680.

\bibitem{raissi2019physics}
M.~Raissi, P.~Perdikaris, and G.~E. Karniadakis, ``{Physics-informed neural networks: A deep learning framework for solving forward and inverse problems involving nonlinear partial differential equations},'' \emph{Journal of Computational Physics}, vol. 378, pp. 686--707, 2019.

\bibitem{10556716}
K.-B. Kwon, R.~Hossain, S.~Mukherjee, K.~Chatterjee, S.~Kundu, S.~Nekkalapu, and M.~Elizondo, ``{Coherency-Aware Learning Control of Inverter-Dominated Grids: A Distributed Risk-Constrained Approach},'' \emph{IEEE Control Systems Letters}, vol.~8, pp. 2120--2125, 2024.

\bibitem{du2023model}
W.~Du~\textit{et al}, ``{Model specification of droop-controlled, grid-forming inverters (REGFM\_A1)},'' Pacific Northwest National Lab, USA, Tech. Rep., 2023.

\bibitem{moya2023approximating}
C.~Moya, G.~Lin, T.~Zhao, and M.~Yue, ``{On approximating the dynamic response of synchronous generators via operator learning: A step towards building deep operator-based power grid simulators},'' \emph{arXiv preprint arXiv:2301.12538}, 2023.

\bibitem{rubanova2019latent}
\BIBentryALTinterwordspacing
Y.~Rubanova, R.~T.~Q. Chen, and D.~Duvenaud, ``{Latent ODEs for irregularly-sampled time series},'' in \emph{Advances in Neural Information Processing Systems}, vol.~32, 2019. [Online]. Available: \url{https://arxiv.org/abs/1907.03907}
\BIBentrySTDinterwordspacing

\bibitem{chen2018neural}
\BIBentryALTinterwordspacing
R.~T.~Q. Chen, Y.~Rubanova, J.~Bettencourt, and D.~Duvenaud, ``{Neural Ordinary Differential Equations},'' in \emph{Advances in Neural Information Processing Systems}, vol.~31, 2018. [Online]. Available: \url{https://arxiv.org/abs/1806.07366}
\BIBentrySTDinterwordspacing

\end{thebibliography}
\bibliographystyle{IEEEtran}
\end{document}